\documentclass[preprint,aps,floats,subeqn,showpacs]{revtex4}
\usepackage{epsfig}

\newcommand{\be}{\begin{equation}}
\newcommand{\ee}{\end{equation}}
\newcommand{\bea}{\begin{eqnarray}}
\newcommand{\eea}{\end{eqnarray}}
\newcommand{\bml}{\begin{mathletters}}
\newcommand{\eml}{\end{mathletters}}

\begin{document}

\tighten

\preprint{IUB-TH-0311}
\draft




\title{Born-Infeld strings in brane worlds}
\renewcommand{\thefootnote}{\fnsymbol{footnote}}
\author{ Yves Brihaye\footnote{Yves.Brihaye@umh.ac.be}}
\affiliation{Facult\'e des Sciences, Universit\'e de Mons-Hainaut,
7000 Mons, Belgium}
\author{Betti Hartmann\footnote{b.hartmann@iu-bremen.de}}
\affiliation{School of Engineering and Sciences, International University Bremen (IUB),
28725 Bremen, Germany}

\date{\today}
\setlength{\footnotesep}{0.5\footnotesep}

\begin{abstract}
We study Born-Infeld strings in a six dimensional brane world scenario recently 
suggested by Giovannini, Meyer and Shaposhnikov (GMS). In the 
limit of the Einstein-Abelian-Higgs model, 
we classify the solutions found by GMS. Especially, we point out that
the warped solutions, which lead to localisation of gravity,
are the - by the presence of the cosmological constant - deformed inverted 
string solutions. Further, we construct the Born-Infeld analogues
of the anti-warped solutions and determine the domain
of existence of these solutions, while a analytic argument leads us to
a ``no-go'' hypothesis: solutions which localise gravity do {\bf not} exist
in a 6 dimensional Einstein-Born-Infeld-Abelian-Higgs (EBIAH) brane world scenario.
This latter hypothesis is confirmed by our numerical results.  
\end{abstract}

\maketitle

\section{Introduction}
The idea that we live in more than the observed $4$ dimensions 
has been of huge interest ever since it was first suggested by Kaluza and Klein
in the 1920s \cite{kk}. They studied a five-dimensional gravitational
theory in a model with an extra {\it compact} dimension. The effective four dimensional
theory then contains $4$ dimensional gravity as well as the electromagnetic
fields and a scalar field. In a similar way, (super)string theories contain
extra {\it compact} dimensions \cite{string} (superstring theories have
$6$ extra compact dimensions) with size of the order of the
Planck length $=1.6 \cdot 10^{-33} cm$. An example is the
Calabi-Yau space in heterotic string theory whose properties
determine the low energy effective field theory.

Other models, which were discussed extensively in recent years are so-called
brane world scenarios \cite{ruba,dvali,anton,arkani,rs1,rs2} which assume that the Standard model (SM) fields
are confined to a $3$-brane (a $3+1$ dimensional submanifold) which is
embedded in a higher dimensional space-time. The extra dimensions
now are {\it non-compact}. Since gravity 
is a property of space-time itself, a model which describes appropriately the
well-tested Newton's law should localise gravity well enough to the $3$-brane.
This was achieved in \cite{rs2} by placing a $3$-brane into $5$ dimensions with the $5$th dimension
being infinite. For the localisation of gravity in this model, the brane tension has to be fine-tuned
to the negative bulk cosmological constant. 

Recently, the localisation of gravity on  different topological defects
has been discussed \cite{shapo}. 
This includes domain walls \cite{dfgk}, Nielsen-Olesen strings \cite{shapo1,shapo2}, 
and magnetic monopoles \cite{shapo3} in $5$, $6$ and 
$7$  space-time dimensions, respectively. It was found \cite{shapo,shapo1}
that gravity-localising (so-called ``warped'') solutions are possible
if certain relations between the defect's tensions hold.
While in the case of domain walls and strings, gravity can only be localised
when the bulk cosmological constant is negative, for magnetic monopoles
the gravity-localisation is possible for both signs of the cosmological constant.

Originally introduced to remove singularities associated
with point-like charges in electrodynamics \cite{BI}, 
the generalisation of the Born-Infeld (BI) action to non-abelian gauge fields has gained a lot
of interest in topics related to string theory \cite{pol,tse}. 
It became apparent that when studying low energy effective actions of
string theory, the part of the Lagrangian containing the abelian Maxwell
field strength tensor and its non-abelian counterpart in Yang-Mills theories
has to be replaced by a corresponding (resp. abelian and non-abelian) BI term.
That's why it seems interesting to generalise the brane world scenario for 6-dimensional
Nielsen-Olesen strings recently proposed by
Giovannini, Meyer and Shaposhnikov (GMS) \cite{shapo2} to Born-Infeld actions.

Our paper is organised as follows : in section II, we present the 6 dimensional
Einstein-Born-Infeld-Abelian-Higgs (EBIAH) model,
we give the equations of motion and especially present our analytic argument
that no solutions, which localise gravity exist in the EBIAH model.
In Section III, we present our numerical results for both the EBIAH model
, namely, we present so-called ``anti-warped'' solutions, as well as
for the limit of the Einstein-Abelian-Higgs (EAH) model. In this latter
case, we put the emphasis on the connection between the GMS solutions
and the four types of solutions available in the EAH model, namely the string branch,
the inverted string branch, the Melvin branch and the Kasner branch \cite{no,gstring}. 
Especially, we show that the ``warped'' solutions found in \cite{shapo1,shapo2} are the - by the cosmological
constant - deformed inverted string solutions. We give our conclusions in Section IV.

\section{The model} 
We have the following $6$-dimensional action \cite{shapo}:
\begin{equation}
S_d=S_{gravity}+S_{brane}
\end{equation}
where the standard gravity action reads
\begin{equation}
S_{gravity}=-\int d^6 x \sqrt{-g} \frac{1}{16\pi G_{6}}\left(R+2\hat\Lambda_6\right) \ .
\end{equation}
$\hat\Lambda_6$ is the bulk cosmological constant, $G_6$ is the fundamental gravity scale with $G_6=1/M^4_{pl(6)}$ and
$g$ the determinant of the $6$-dimensional metric.

The action $S_{brane}$ for the 
Einstein-Born-Infeld-Abelian-Higgs (EBIAH) string is given by \cite{no,yves}:

\begin{equation}
S_{brane}=\int d^6 x \sqrt{-g_6} \left(\hat{\beta}^2(1-{\cal R})+\frac{1}{2}D_M\phi D^M\phi^*-\frac{\lambda}{4}(\phi^*\phi-v^2)^2  \right)
\end{equation}
where 
\begin{equation}
{\cal R}=\sqrt{1+\frac{F_{MN}F^{MN}}{2\hat{\beta}^2}-\frac{(F_{MN}\tilde{F}^{MN})^2}{16\hat{\beta}^4}}
\end{equation}
with the covariant derivative $D_M=\nabla_M-ieA_M$ and the field strength $F_{MN}=\partial_M 
A_N-\partial_N A_M$ of the U(1) gauge potential $A_M$.
$v$ is the vacuum expectation value of the complex valued Higgs field $\phi$ and
$\lambda$ is the self-coupling constant of the Higgs field. 

\subsection{The Ansatz}
The Ansatz for the $6$-dimensional metric reads \cite{shapo}:
\begin{equation}
ds^2= M^2(r)\left(dx_1^2-dx_2^2-dx_3^2-dx_4^2\right)-dr^2- l^2(r)d\theta^2  \ .
\end{equation}
For the gauge and Higgs field, we have \cite{no}:
\begin{equation}
\phi(r, \theta)=v f(r) e^{i n\theta} \ , \ \ A_{\theta}(r,\theta)=\frac{1}{e}(n-P(r))
\end{equation}
where $n$ is the vorticity of the string.
\subsection{Equations of Motion}
Introducing the following dimensionless coordinate $x$ and the dimensionless function $L$:
\begin{equation}
x=\sqrt{\lambda}v r \ , \ \ \ L(x)=\sqrt{\lambda}v l(r)
\end{equation}
the set of equations depends only on the following dimensionless coupling constants:
\begin{equation}
\alpha=\frac{e^2}{\lambda} \ , \ \ \ \gamma^2=8\pi G_6 v^2 \ , \ \ \ 
\Lambda=\frac{\hat\Lambda_6}{\lambda v^2} \ , \ \ \ 
\beta^2=\frac{\hat{\beta}^2}{\lambda v^4} \ , \ \ 
\end{equation}
The gravitational equations then read:
\begin{equation}
\label{eq1}
3 \frac{M^{''}}{M} + \frac{L^{''}}{L}
+ 3 \frac{L^{'}}{L} \frac{M^{'}}{M} 
+ 3 \frac{M^{'2}}{M^2}
+\Lambda  =-\gamma^2 T^0_0
\end{equation}

\begin{equation}
6 \frac{M^{' 2}}{M^2} + 
4 \frac{L^{'}}{L} \frac{M^{'}}{M} +\Lambda
= -\gamma^2 T^x_x \ ,
\end{equation}

\begin{equation}
\label{eq3}
6 \frac{M^{' 2}}{M^2}+
4 \frac{M^{''}}{M} +\Lambda =
-\gamma^2 T^{\theta}_{\theta} \ ,
\end{equation}
where
\begin{equation}
\label{t0}
T^0_0=\left(\frac{(f^{'})^2}{2}+\frac{(1-f^2)^2}{4}+\frac{f^2 P^2}{2L^2}
+\beta^2\left(\sqrt{1+\frac{P^{' 2}}{\alpha\beta^2 L^2}}-1\right)\right) \ ,
\end{equation}
\begin{equation}
T^{x}_{x}=\left(-\frac{(f^{'})^2}{2}+\frac{(1-f^2)^2}{4}
+\frac{f^2 P^2}{2L^2} +\beta^2\left(\sqrt{1+\frac{P^{' 2}}{\alpha\beta^2 L^2}}-1\right)-\frac{P^{' 2}}{\alpha L^2}
\left(\sqrt{1+\frac{P^{' 2}}{\alpha\beta^2 L^2}}\right)^{-1}\right)
\end{equation}
\begin{equation}
\label{ttheta}
T^{\theta}_{\theta}=\left(\frac{(f^{'})^2}{2}+\frac{(1-f^2)^2}{4}-\frac{f^2 P^2}{2L^2}
+\beta^2\left(\sqrt{1+\frac{P^{' 2}}{\alpha\beta^2 L^2}}-1\right)-\frac{P^{' 2}}{\alpha L^2}
\left(\sqrt{1+\frac{P^{' 2}}{\alpha\beta^2 L^2}}\right)^{-1}\right) \ .
\end{equation}
The Euler-Lagrange equations for the matter fields read:
\begin{equation}
\frac{(M^4 L f^{'})^{'}}{M^4L}+
(1-f^2)f-\frac{P^2}{L^2}f=0
\end{equation}
and
\begin{equation}
\label{peq}
\label{eqp}
\frac{L}{M^4}\left(\frac{M^4 P^{'}}{L \sqrt{1+\frac{P^{' 2}}{\alpha\beta^2 L^2}}}\right)^{'}-\alpha f^2 P=0 \ .
\end{equation}
The prime denotes the derivative with respect to $x$. 
For $\beta^2=\infty$ these equations reduce to those studied in \cite{shapo}.

The equations (\ref{eq1})-(\ref{eq3}) can be combined to
obtain the following two differential equations for the two unknown metric functions:
\begin{equation}
\frac{(M^4 L^{'})^{'}}{M^4 L}+\frac{\Lambda}{2}
=-\gamma^2\left(T_0^0+\frac{1}{4} T_x^x-\frac{3}{4}T_{\theta}^{\theta}\right)
\end{equation}
which then gives:
\begin{equation}
\frac{(M^4 L^{'})^{'}}{M^4 L}+\frac{\Lambda}{2}=-\gamma^2\left(\frac{1}{8}
(1-f^2)^2+\frac{f^2 P^2}{L^2}+\frac{\beta^2}{2}
\left(\sqrt{1+\frac{P'^2}{\alpha\beta^2 L^2}}-1\right)
+\frac{P'^2}{2\alpha L^2}\left(\sqrt{1+\frac{P'^2}{\alpha\beta^2 L^2}}\right)^{-1}\right)
\end{equation}
and
\begin{equation}
\frac{(LM^3 M^{'})^{'}}{M^4 L}+\frac{\Lambda}{2}=-\frac{\gamma^2}{4}\left(T_x^x+
T_{\theta}^{\theta}\right)
\end{equation}
which then reads:
\begin{equation}
\frac{(LM^3 M^{'})^{'}}{M^4 L}+\frac{\Lambda}{2}
=-\frac{\gamma^2}{4}\left(\frac{(1-f^2)^2}{2}+2\beta^2\left(\sqrt{1+\frac{P'^2}
{\alpha\beta^2 L^2}}-1\right)-\frac{2P'^2}{\alpha L^2}\left(\sqrt{1+\frac{P'^2}{\alpha\beta^2 L^2}}\right)^{-1}  \right)
\end{equation}
\subsection{Boundary conditions}
We require
regularity at the origin $x=0$ which leads to the boundary conditions:
\begin{equation}
\label{bcx0}
f(0)=0 \ , \ \ \ P(0)=n \ , \ \ \ M(0)=1 \ , \ \ \ M^{'}|_{x=0}=0 \ , \ \ \
L(0)=0 \ , \ \ \  L^{'}|_{x=0}=1 \ .
\end{equation}
The requirement for finiteness of the energy leads to:
\begin{equation}
f(\infty)=1 \ , \ \ \ P(\infty)=0 \ .
\end{equation}

\subsection{String tensions}
The string tensions are defined as follows:
\begin{equation}
\rho_i=\int_0^{\infty} dx \sqrt{-g} T_i^i \ .
\end{equation}
The relation between the tensions $\rho_0$ and $\rho_{\theta}$ can then 
be evaluated. Using (\ref{t0}) and (\ref{ttheta}), we find:
\begin{equation}
\rho_0-\rho_{\theta}=\int_0^{\infty} dx M^4 L \left(\frac{f^2 P^2}{L^2}+
\frac{P^{'2}}{\alpha L^2 \sqrt{1+\frac{P^{' 2}}{\alpha\beta^2 L^2}}}\right)
\end{equation}
and using (\ref{eqp}), we find:
\begin{equation}
\rho_0-\rho_{\theta}=\frac{1}{\alpha}\left(\frac{M^4 P P^{'}}
{\sqrt{L^2+\frac{P^{' 2}}{\alpha\beta^2}}}\right)\Biggl\vert^{\infty}_{0} \ .
\end{equation}
For warped solutions, the infinity term should vanish, while
for $x=0$, we use the boundary conditions (\ref{bcx0}). Thus:
\begin{equation}
\label{rho}
\rho_0-\rho_{\theta}=-\frac{n |\beta|}{\sqrt{\alpha}}
\end{equation}
Now, from \cite{shapo1}, we know that $\rho_0-\rho_{\theta}=\gamma^{-2}$ in order
to have solutions which localise gravity. However, since the rhs of (\ref{rho})
is always negative, this relation can never be fulfilled, which leads us to
the hypothesis that warped solutions don't exist in the Einstein-Born-Infeld model.

\section{Numerical solutions}
Before to discuss our numerical results for the EBIAH model,
let us explain the pattern of solutions in the $\beta=\infty$ case i.e. in the
Einstein-Abelian-Higgs (EAH) model. 
Since it was stated in \cite{shapo2} that no attempt to classify the 
solutions is done, the following discussion can be seen as such a classification.

\subsection{Einstein-Abelian-Higgs (EAH) model ($\beta=\infty$ limit)}
We first discuss the solutions in the case $\Lambda=0$.
The pattern of solution in the 6-dimensional EAH model 
is very similar to the one in the 4-dimensional EAH model \cite{gstring}~:
for generic values of $\alpha$, $\gamma$ there are two branches of
solutions: a branch for which $M(x \rightarrow \infty)=a$
and $L(x\rightarrow \infty) = b x + c$, where the parameters
$a$, $b$ depend on the coupling constants. For $\gamma=0$,
the Nielsen-Olesen string solutions \cite{no} are recovered,
this is why this branch of solutions is referred to as the {\it string branch}.

For $\gamma \ll 1$ the Nielsen-Olesen string
is slightly deformed by gravity and $b \approx 1 $ , $a \approx 1$.
When $\gamma$ is increased (with $\alpha$ being fixed) the
parameter $a$ varies slowly while the parameter $b$ decreases linearly
with $\gamma$ such that it becomes negative at some critical value
of $\gamma$, say $\gamma = \gamma^0_{cr} (\alpha,\Lambda=0)$. For
$\gamma > \gamma^0_{cr}$ the solutions continue to exist but, since the
function $L(x)$ possesses a zero , the solution is not regular on the full
interval of the radial coordinate.
The solutions for $\gamma > \gamma^0_{cr}$ are called {\it inverted 
string} solutions. We have constructed solutions for the self-dual limit ($\alpha=2$)
and for $\Lambda=0$. We find that in complete agreement with the
four dimensional case, the critical value $\gamma^0_{cr}(\alpha=2)=2$ and
that in addition $M^{\alpha=2}(x)=1$, while $b^{\alpha=2}
(\gamma)=1-\frac{\gamma}{2}$ (see Fig.~1). Thus for $\gamma > \gamma^0_{cr}=2$, 
$L(x)$ is zero
at some $x=x_0(\gamma)$, $L(x=x_0)=0$. Note that the function $M(x)$ stays finite
at $x=x_0$.

For the same values of the coupling constants
a second branch of solutions, the so-called {\it  Melvin branch}, exists.
In $d=6$, the solutions on this branch behave like
$M(x\rightarrow \infty) = A x^{2/5}$, $L(x\rightarrow \infty) = B
x^{-3/5}$. The constants $A$ and $B$ depend on the coupling constants
of the model. There exist also solutions, the so-called {\it Kasner solutions}, 
for $\gamma > \gamma^0_{cr}$. For these strongly coupled gravitating solution
the function $M(x)$ develops  a node at some finite value
of $x=\tilde{x}_0$ with $M(x=\tilde{x}_0)=0$, while  
at the same time the function $L(x\rightarrow \tilde{x}_0)\rightarrow \infty$.
For $\alpha=2$, we present the dependence of the values $A$ and $B$ on the
gravitational coupling $\gamma$ in Fig.~1. We find that like in the case of the string
branch $\gamma^0_{cr}(\alpha=2)=2$. For $\gamma\rightarrow 2$, $A$ tends to
zero, while $B$ tends to infinity.

From the discussion of the solutions in the $\Lambda=0$ limit, we note that the main
feature is the existence of a critical value of $\gamma$, say $\gamma^0_{cr}(\alpha,\Lambda=0)$,
which clearly separates the classical solution into two distinguished domains;
string branch and Melvin branch exist for $\gamma < \gamma^0_{cr}(\alpha,0)$,
inverted string branch and Kasner branch for $\gamma > \gamma^0_{cr}(\alpha,0)$.
 However, as was noted
in \cite{gstring}, the domain of existence of the Kasner solutions
is limited. For the 4-dimensional Kasner solutions, it was found
that for fixed $\gamma^{d=4}=1.0$ and varying $\alpha^{d=4}$, the value
$\tilde{x}_0^{d=4}$ tends to infinity at $\alpha^{d=4}\approx 0.15$.
Thus, we expect that in analogy in $d=6$, the Kasner solutions
will exits only for small $\alpha$ when $\Lambda=0$. Especially, for $\alpha=2.0$ (which
is the value of $\alpha$ mainly considered in this paper), we expect
to find no Kasner solutions for $\Lambda=0$, $\gamma > 2$. This is confirmed
by our numerical results. 

We will now discuss, how this pattern evolves when the cosmological constant
is negative and construct the domain of existence of solutions 
in the $\gamma$-$\Lambda$-plane (with $\alpha$ fixed and $\gamma \geq 0$, $\Lambda\leq 0$). Our result
are summarised in Fig.~2 for $\alpha=2$, but we expect that the
results are similar for generic values of $\alpha$.

Considering first the part of the plane for which $\gamma < 2$, 
our numerical results indicate that both string branch and Melvin branch exist
up to a critical value $\gamma < \tilde{\gamma}_{cr}(\alpha, \Lambda)$ (or equivalently
$\Lambda > \tilde{\Lambda}_{cr}(\alpha,\gamma)$). In the limit
$\gamma \rightarrow \tilde{\gamma}_{cr}(\alpha,\Lambda)$, the two solutions converge to a common
solution indicated by the dashed line in Fig.~2. This phenomenon
persists also for $\beta$ finite and will thus be discussed in more detail in the next section.
In order to make the connection with \cite{shapo2}, 
let us mention that the so-called ``anti-warped'' solutions in Fig.s 15 and 17 of \cite{shapo2} 
belong, respectively, to the string and Melvin branch.

Let us now discuss the part of the $\gamma$-$\Lambda$-plane with $\gamma > 2$.
As indicated in Fig.~2, below a critical line (solid), which we denote 
$\Lambda_{cr}(\alpha, \gamma)$,
only inverted string solutions exist, while above this line only Kasner
solutions exist.
The critical phenomenon which limits the domain of existence of the 
inverted string solutions
is related to the occurence of a zero $x=x_0$ of the function $L(x)$. Indeed,
the numerical study of $x_0$  
for $\alpha$, $\gamma$ fixed and varying $\Lambda$, reveals that $x_0$ 
is shifted
to infinity for a finite value of $\Lambda$, which coincides with the
critical value of $\Lambda$, $\Lambda_{cr}(\alpha, \gamma)$, mentioned above.
This phenomenon is illustrated in Fig.~3 for two pairs of
values ($\alpha$, $\gamma$) such that $\gamma > \gamma^0_{cr}(\alpha,\Lambda=0)$. We plot
the quantity $x_0$ in dependence on the bulk cosmological
constant $\Lambda$. Clearly at some $\Lambda=\Lambda_{cr}(\alpha,\gamma)$ $x_0$
tends to infinity 
and it can be checked 
that the functions $L(x)$ and $M(x)$ become exponentially decaying.
We find that $\Lambda_{cr}(\alpha=1,\gamma=1.75)\approx -0.0014$ and
 $\Lambda_{cr}(\alpha=2,\gamma=2.157)\approx -0.0035$, which is in perfect
agreement with the results of \cite{shapo1,shapo2}.
Stated in another way, the crucial point about the construction in \cite{shapo2} 
is that the cosmological constant has to be
fine tuned in such a way that the zero of the function $L(x)$
is pushed to infinity and consequently the metric functions become exponentially
decaying:
\begin{equation}
M(x\rightarrow \infty)=M_0 \exp(-\sqrt{-\Lambda/10}x)
 \ \ , \ \ L(x\rightarrow \infty)=L_0 \exp(-\sqrt{-\Lambda/10}x) \ \ .
\end{equation}

\subsection{Einstein-Born-Infeld-Abelian-Higgs (EBIAH) model ($\beta < \infty$)}
Classifying the solutions of the full EBIAH equations 
is a vast task because of the occurence of four parameters to be varied.
First, we therefore limited our investigation to the case $\alpha=1$ and $\beta^2=5$.
Along with the case $\beta=\infty$ discussed above, we orientated
first our numerical construction to small value of $\gamma$. In this case,
the string and Melvin branches continue to exist and the phenomenon
that these two branches converge to a common solution for a $\Lambda$-depending
critical value of $\gamma$ is recovered. This is illustrated in
Fig.~4, where we plot the metric functions $L(x)$ and $M(x)$ of 
the ``anti-warped'' solutions corresponding to the string
and Melvin branches for $\Lambda=-0.1$. In this case, the critical
value of $\gamma$ is $\tilde{\gamma}_{cr}
(\alpha=1, \beta=\sqrt{5}, \Lambda=-0.1)\approx 1.0625$.

With this information, we constructed the domain of existence of solutions
in the $\gamma$-$\Lambda$-plane for $\alpha=2$ and $\beta=1$.
First, we determined the critical value of $\gamma$, 
$\gamma^0_{cr}(\alpha=2,\beta=1,\Lambda=0)$,
for which the string and Melvin solutions turn into closed solutions.
We find that $\gamma^0_{cr}(\alpha=2,\beta=1,\Lambda=0)\approx 2.045$
which differs only little from the value in the $\beta=\infty$ limit.
For $\gamma < \gamma^0_{cr}$, we find -as mentioned before- that
the domain of existence of the string and Melvin solutions is
restricted by a curve $\tilde{\gamma}_{cr}(\alpha=2,\beta=1,\Lambda)$ (or
$\tilde{\Lambda}_{cr}(\alpha=2,\beta=1,\gamma)$), at which the two solutions
converge to a common solution (dashed line in Fig.~5). This is in complete
analogy to the $\beta=\infty$ limit. In this part of the $\gamma$-$\Lambda$-plane, however,
the domain of existence of the Melvin solutions is restricted by another 
phenomenon. At some $\tilde{\gamma}^M_{cr}(\alpha=2,\beta=2,\Lambda)$
(or $\tilde{\Lambda}^M_{cr}(\alpha=2,\beta=1,\gamma)$, dotted-dashed line
in Fig.~5) with 
$\tilde{\gamma}^M_{cr} < \tilde{\gamma}_{cr}$, the value of the magnetic
field at the origin tends to infinity. This phenomenon was previously noted
for the $d=4$, $\Lambda=0$ case in \cite{yves}. Thus, Melvin
solutions exist only in a small part of the $\gamma$-$\Lambda$-plane, 
namely for $\tilde{\gamma}^M_{cr} < \gamma < \tilde{\gamma}_{cr}$.

Now turning to solutions for $\gamma > \gamma^0_{cr}$, again, we find
no Kasner solutions for $\Lambda=0$ since $\alpha=2$ is too large.
We were only able to construct inverted string solutions for $\alpha=2$, $\beta=1$ and
$\Lambda=0$. The inverted string solutions exist up to a critical
value $\Lambda_{cr}(\alpha=2,\beta=1,\gamma)$ (or 
$\gamma_{cr}(\alpha=2,\beta=1,\Lambda$), solid line
in Fig.~5). For $\Lambda > \Lambda_{cr}(\alpha=2,\beta=1,\gamma)$ only
Kasner solutions exist. Following our analytic argument in Section II.D.
no ``warped solutions'' exist such that the solid line in Fig.~5 now represents
a line where no solutions at all exist. Thus our numerical results
confirm the analytic argument that no gravitation-localising
solutions exist in a Born-Infeld brane world scenario.  

\section{Conclusions}
In this paper, we have studied a 6-dimensional Einstein-Born-Infeld-Abelian-Higgs
model in a brane world scenario recently suggested by \cite{shapo2}.
In the $\beta=\infty$ limit, our model reduces to that of \cite{shapo2}.
As a cross-check of our model, we reconsidered the solutions found in 
this paper, and managed to determine the domain of the $\gamma$-$\Lambda$-plane
in which solutions of this model exist. Especially, we identify the
four different types of branches that exist for $\Lambda=0$, namely the string, Melvin, inverted
string and Kasner branches and put the emphasis on the extention of
these branches in the $\Lambda < 0 $ case. We identified the warped solution
of \cite{shapo2} as the solution sitting on the limit of the domain of
existence of the inverted string solution (decreasing $\Lambda$ and keeping
the other parameters fixed). We also considered the influence of the
Born-Infeld interaction on the solutions and recovered the existence
of the four types of solutions in this case. 
We construct the domain of existence of solutions in the $\gamma$-$\Lambda$-plane
for $\alpha=2$ and $\beta=1$. Especially, we find that in agreement 
with our analytic
argument, which uses the reasoning based on the string tensions of 
\cite{shapo2}, no ``warped solutions'' exist in the Born-Infeld case.

Let us remark that we have only taken into account the
Born-Infeld terms of the effective action of string theory and have
neglected higher derivative corrections in this paper. These arise
naturally in the effective action of open string theory and are related
to the essential non-locality of string field theory. These terms could,
of course, influence the ``no-go'' result of this paper. We plan to
address this question in a future publication.

 \newpage
\begin{figure}
\centering
\epsfysize=12cm
\mbox{\epsffile{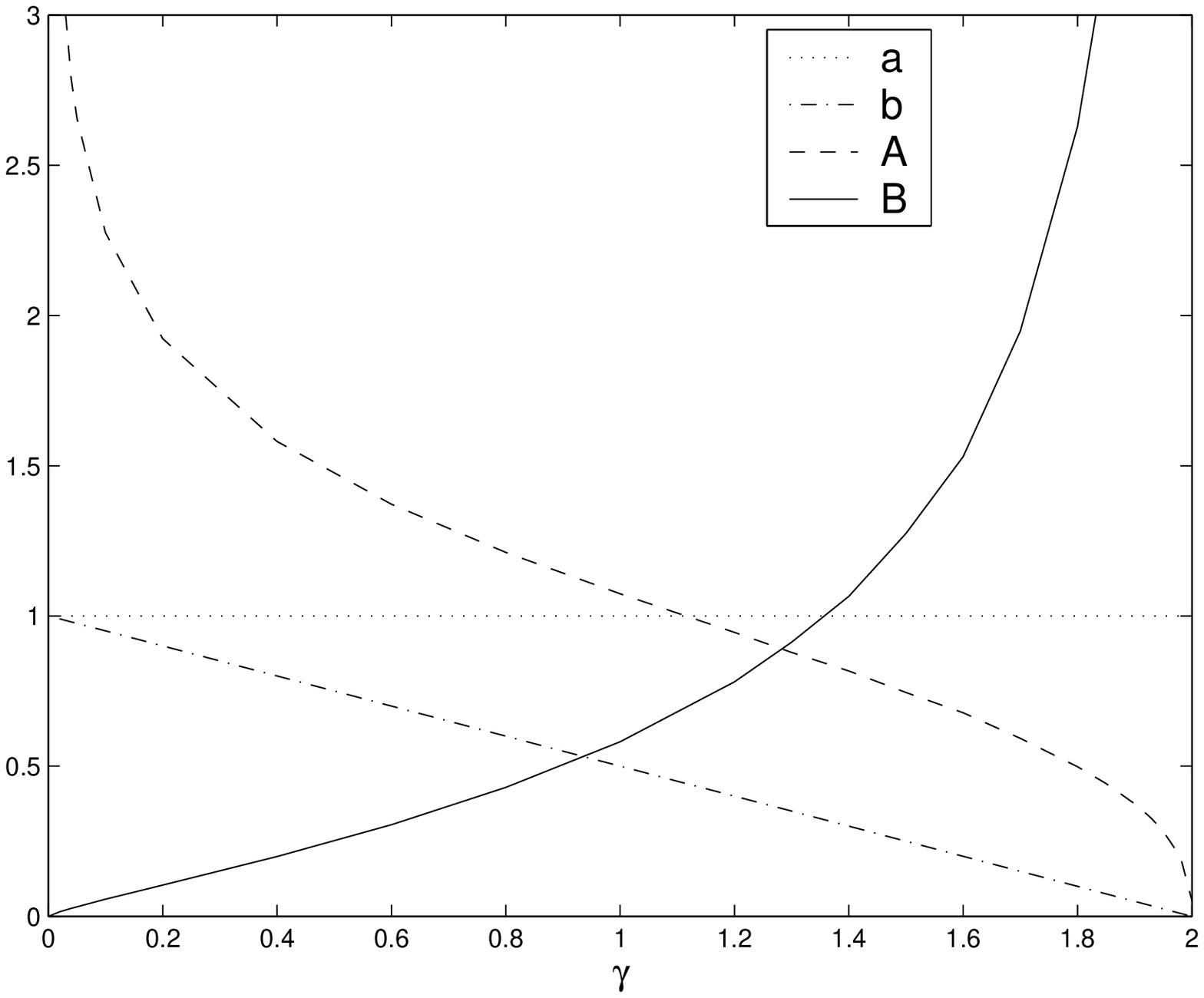}}
\caption{\label{Fig.1} The dependence of the parameters $A$, $B$, $a$ and $b$
on $\gamma$ is shown for the solutions in the $\beta=\infty$, $\alpha=2$ limit
on the string branch ($M(x\rightarrow\infty)=a$,
$L(x\rightarrow\infty)=bx+c$) and on the Melvin branch ($M(x\rightarrow \infty)=
Ax^{2/5}$, $L(x\rightarrow\infty)=Bx^{-3/5}$), respectively. }
\end{figure}

\begin{figure}
\centering
\epsfysize=12cm
\mbox{\epsffile{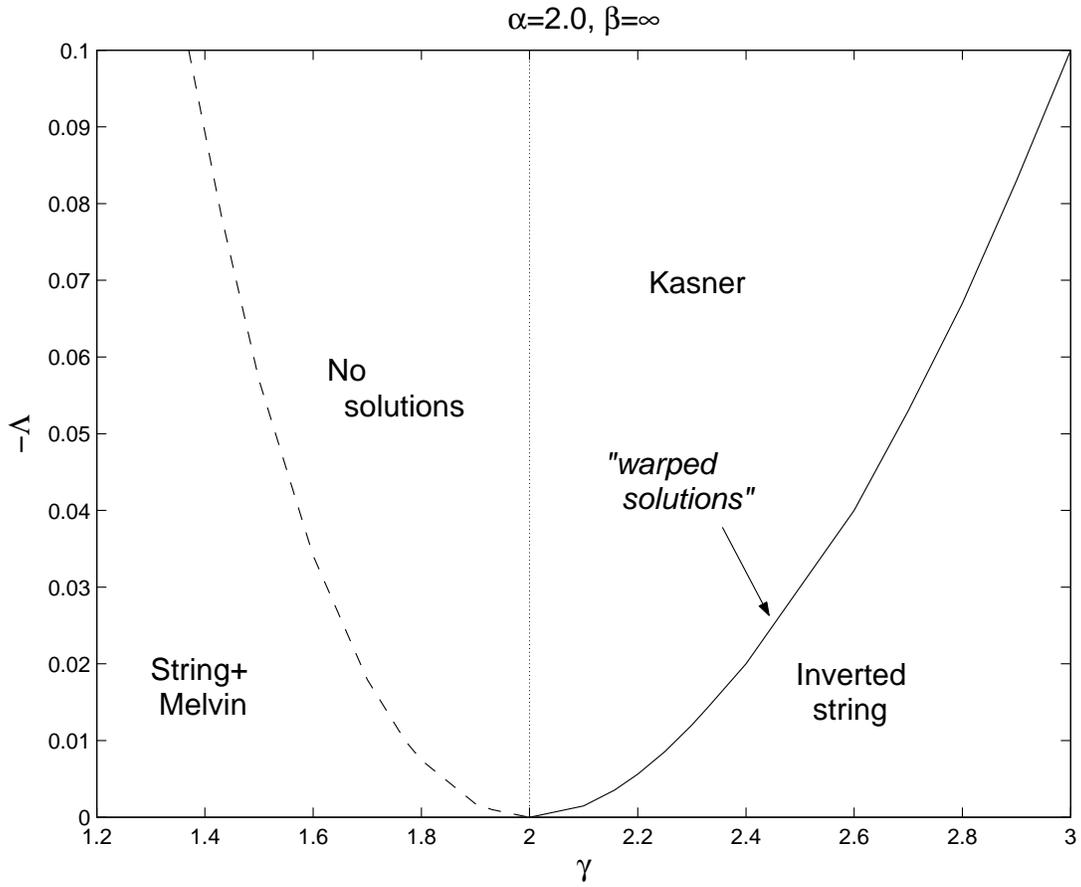}}
\caption{\label{Fig.2} The domain of existence of solutions in the $\gamma$-$\Lambda$-plane
is presented for $\alpha=2.0$, $\beta=\infty$. Note that the dashed line represents $\tilde{\gamma}_{cr}(2,\Lambda)$, while
the solid line denotes $\Lambda_{cr}(2,\gamma)$.  }
\end{figure}

\begin{figure}
\centering
\epsfysize=12cm
\mbox{\epsffile{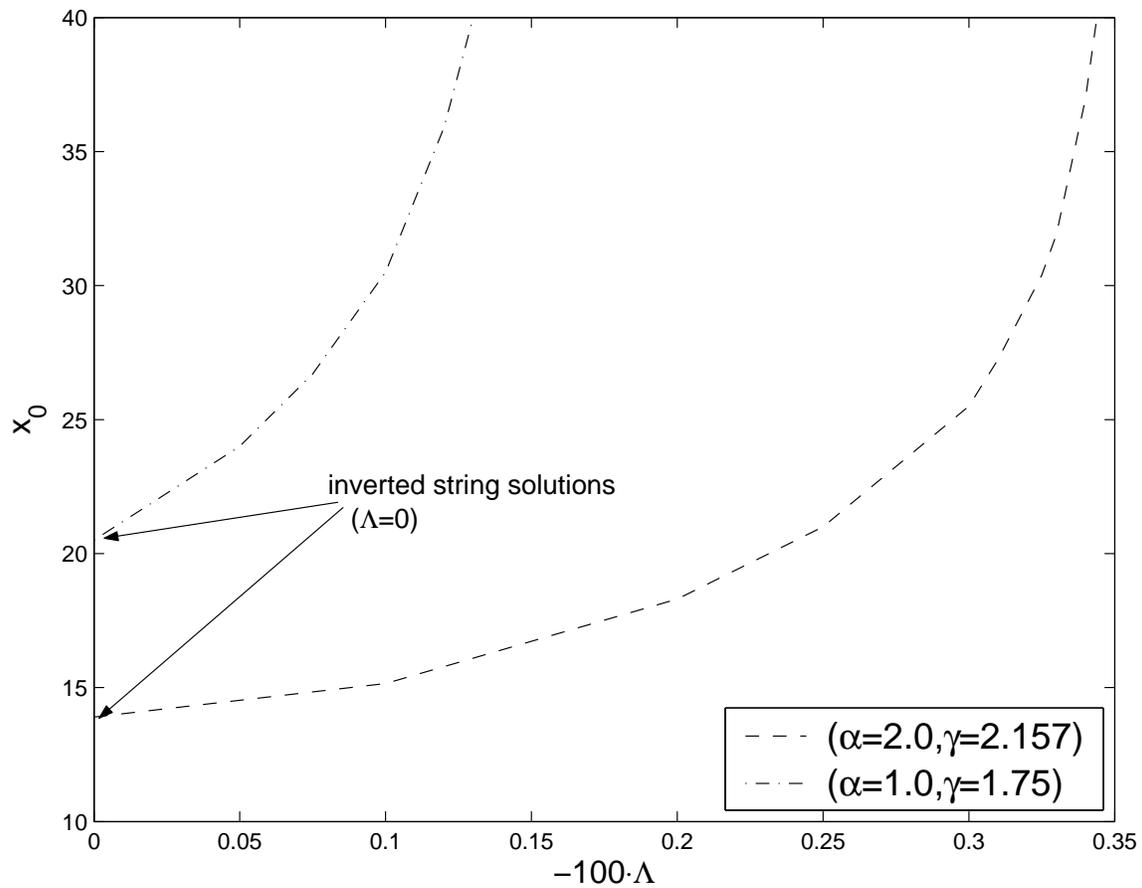}}
\caption{\label{Fig.3} The value of the quantity $x_0$ ($L(x=x_0)=0$) is plotted
as function of $\Lambda$ \
for two different pairs of values $(\alpha,\gamma)$ and $\beta=\infty$. }
\end{figure}

\begin{figure}
\centering
\epsfysize=15cm
\mbox{\epsffile{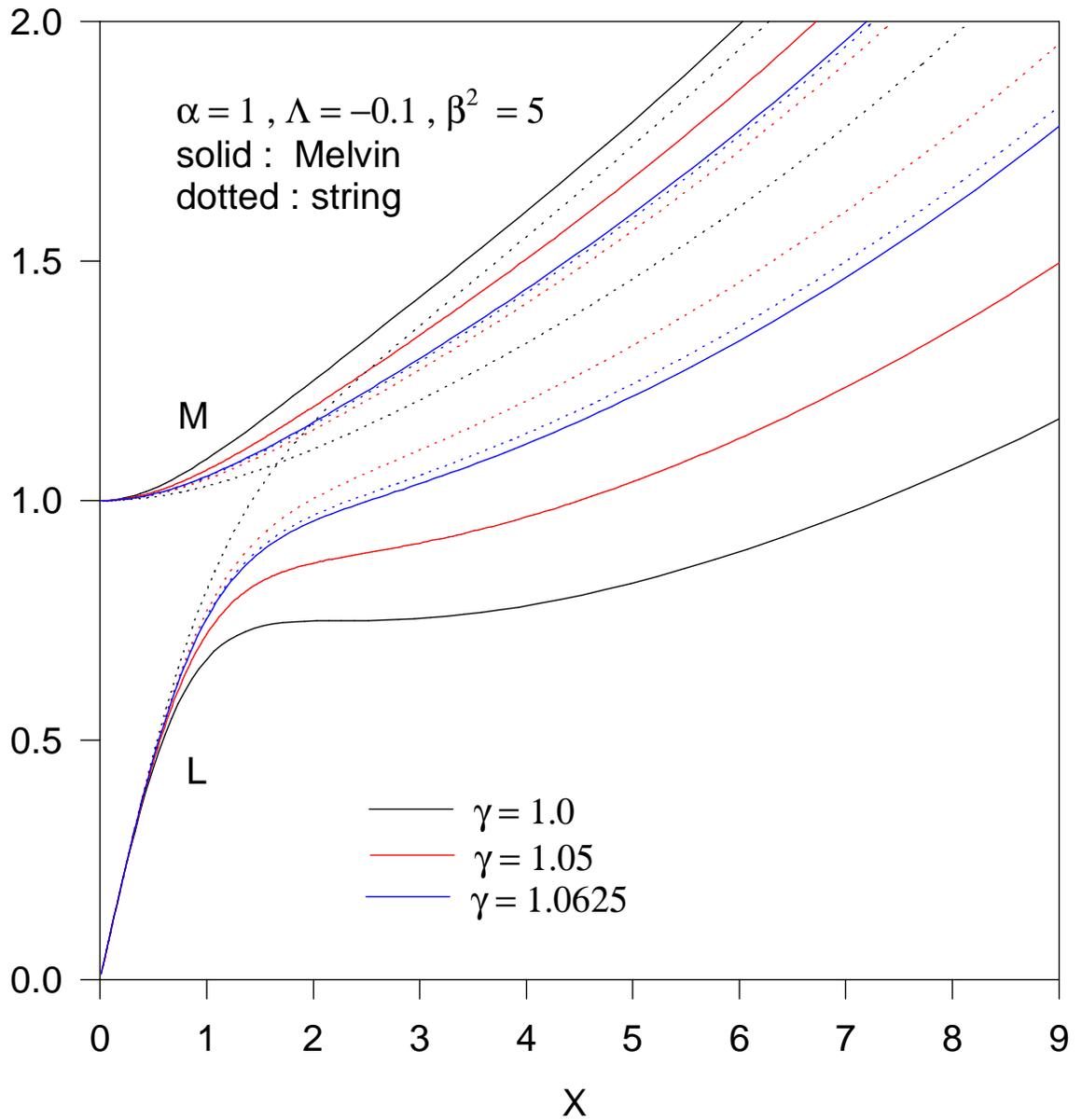}}
\caption{\label{Fig.4} {\it (Colour figure)} The ``anti-warped'' solutions of the Born-Infeld 
equations with $\alpha=1.0$, $\Lambda=-0.1$, $\beta^2=5.0$  are presented
for different values of $\gamma$ ($\gamma=1.0$ (black), $\gamma=1.05$ (red)
and $\gamma=1.0625$ (blue)) approaching 
$\gamma_{cr}(\alpha=1,\beta^2=5,\Lambda=-0.1)\approx 1.063$. }
\end{figure}

\begin{figure}
\centering
\epsfysize=12cm
\mbox{\epsffile{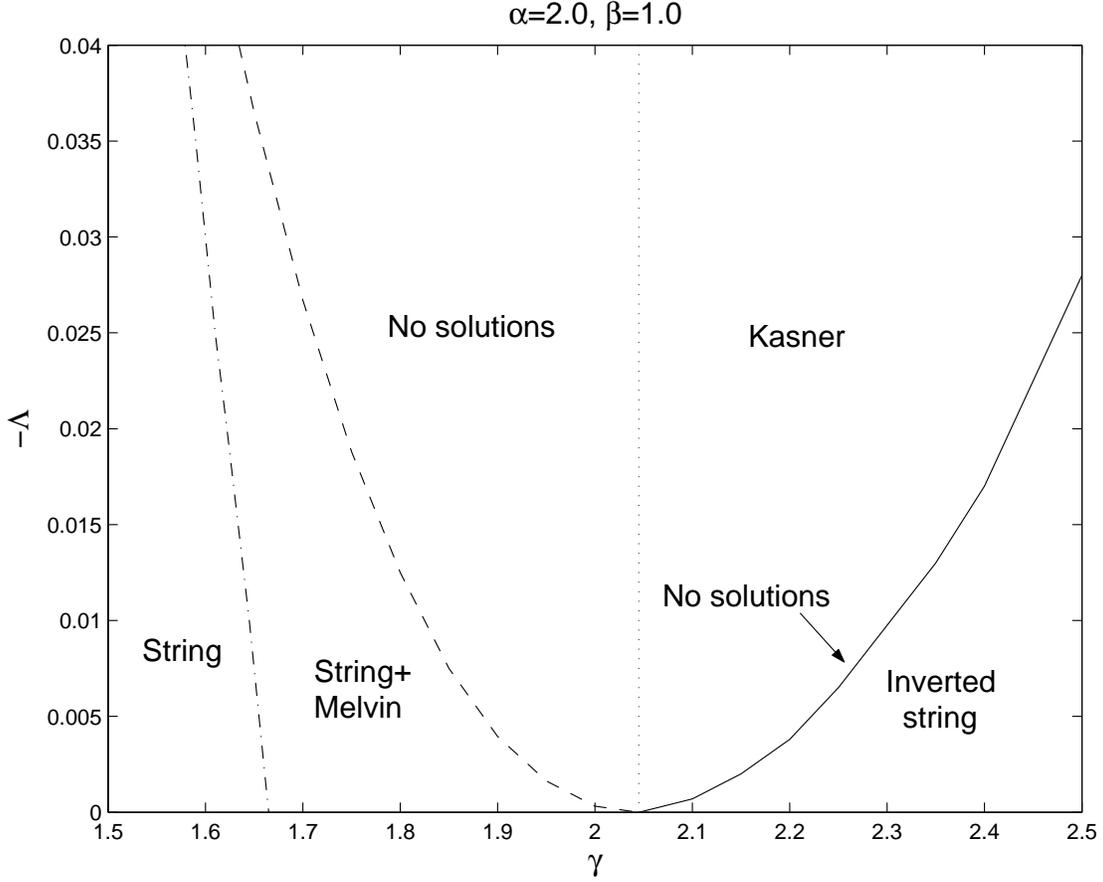}}
\caption{\label{Fig.5} The domain of existence of solutions in the $\gamma$-$\Lambda$-plane
is presented for $\alpha=2.0$, $\beta=1.0$. 
Note that the dotted-dashed line corresponds to those values of
$\gamma$ and $\Lambda$ for which the magnetic field of the Melvin
solutions tends to infinity at the origin ($\tilde{\gamma}^M_{cr}(\alpha=2,\beta=1,
\Lambda)$).
The dashed line corresponds to $\tilde{\gamma}_{cr}(\alpha=2,\beta=1,\Lambda)$
at which the string and Melvin solutions converge to a common solution,
while the solid line represents the limit of existence of inverted string solutions,
$\Lambda_{cr}(\alpha=2,\beta=1,\gamma)$. The dotted line corresponds to $\gamma^0_{cr}(\alpha,\beta,\Lambda=0)$. 
  }
\end{figure}

\end{document}